\documentstyle[seceq,preprint,mbf,wrapft,epsf]{ptptex}

\def\P{\mathop{\mathrm{{{\mbf P}}}^{(i)}}}
\newcommand{\Slash}[1]{\ooalign{\hfil/\hfil\crcr$#1$}}

\notypesetlogo  

\title{Properties of Three-Body Decay Functions Derived with Time-Like Jet Calculus beyond Leading Order}

\author{Tetsuya {\sc Sugiura}\footnote{E-mail: sugiura@fmr.rikkyo.ne.jp}}

\inst{Department of Physics, Rikkyo University, Tokyo 171-8501, Japan}

\recdate{ December 7,  2001}

\abst{Three-body decay functions in time-like parton branching are calculated using the jet calculus to the next-to-leading logarithmic (NLL) order in perturbative quantum chromodynamics (QCD). The phase space contributions from each of the ladder diagrams and interference diagrams are presented. We correct part of the results for the three-body decay functions calculated previously by two groups. Employing our new results, the properties of the three-body decay functions in the regions of soft partons are examined numerically. Furthermore, we examine the contribution of the three-body decay functions modified by the restriction resulting from the kinematical boundary of the phase space for two-body decay in the parton shower model. This restriction leads to some problems for the parton shower model. For this reason, we propose a new restriction introduced by the kinematical boundary of the phase space for two-body decay.}

\begin{document}

\maketitle

\makeatletter
\if 0\@prtstyle
\def\asp{0.5em} \def\bsp{0em}
\else
\def\asp{.3em} \def\bsp{0.3em}
\fi \makeatother

\section{Introduction}
 The next generation of linear colliders (NLC), such as the Japan linear collider (JLC)\cite{rf:1} and the super conducting electron positron linear collider (TESLA)\cite{rf:2} will be running in the near future. The energies of these accelerators will be in the TeV region. Many multijets will be produced in this energy region. In order to analyze the experimental data for such multijet events, parton shower models based on perturbative QCD will be useful. To this time, many parton shower models limited to the contribution of the leading logarithmic (LL) order have been constructed.\cite{rf:3} 
 
 With the purpose of constructing a parton shower model including the contribution of the next-to-leading logarithmic order, three-body decay functions have been calculated previously by two groups.\cite{rf:4,rf:5} Using these results, NLLJET has been constructed, and it has been used to analyze experimental data, such as that from SLAC-PEP, TRISTAN and LEP1. \cite{rf:6}
 
 At present, only two groups have calculated the three-body decay functions. Since some discrepancies have been found in the results reported in Refs.~\citen{rf:4} and \citen{rf:5}, the results for the three-body decay functions must be verified. 
 
 The three-body decay functions in the parton shower model have been modified with respect to the manner in which the effect of the one soft gluon is absorbed into the kinematical boundary in the phase space of the two-body decay.\cite{rf:4,rf:9} This modification corresponds to the angular ordering of the three-body decay functions.
 
 We expect that many soft partons will be produced in the TeV energy region. In order to decrease the systematic error involved in the data analyses of multijet events, the effects of the soft partons should be included in the parton shower model. For example, there is the double soft gluon limit. Theoretical studies of the soft parton for the three-body decay functions are necessary to improve the parton shower model, such as NLLJET.
 
 In this paper, we employ the method of Ref.~\citen{rf:7} to construct the kinematics of three-body decay functions in time-like branching. The contributions of the three-body decay functions for the soft partons are then examined using the results so obtained.
 
 The organization of the paper is as follows. In the next section, we construct the phase volume element. The details of the kinematics are given in Appendix A. In $\S$3, the calculated results are presented. We find that part of the results obtained in this paper do not agree with those obtained previously by two groups. In $\S$4, the properties of the calculated three-body decay functions are examined. In $\S$5, we give some comments on the angular ordering. In the final section, we give a summary of our conclusions.

\section{Kinematics for the calculation}
 In this section, the kinematics used to calculate the three-body decay functions are presented. The three-body decay functions are determined from the time-like jet calculus on the basis of the QCD factorization theorem.\cite{rf:4,rf:5,rf:10} To begin with, the factorization is introduced following Ref.~\citen{rf:10}. In the light-cone gauge, collinear singularities appear from the on-shell partons (parton internal lines) between two-particle irreducible (2PI) kernels, and they are factorized using the projection operator. 
 
 We use $A$ and $B$ to represent the 2PI kernels that are connected by gluon internal lines or quark internal lines. The amplitude squared $C$ is defined as

\begin{eqnarray}C=A_{\mu'\nu'}B^{\mu'\nu'}\end{eqnarray}
for gluon internal lines and 

\begin{eqnarray}C={\mathrm{Tr}}[A B]\end{eqnarray}
for quark internal lines, where $\mu'$ and $\nu'$ are Lorentz indices. The projection operators to extract the collinear singularities are obtained from the general formulas of the 2PI kernels. \cite{rf:10} We use $q$ to represent momenta on the internal lines that correspond to the on-shell partons. Inserting the projection operator $P^{(G)}(q)$ for gluon internal lines into Eq. (2$\cdot$1), we have

\begin{eqnarray}P^{(G)}(q)C=\Bigl[A_{\mu'\nu'}d^{\mu'\nu'}(q)\Bigr]_{q^2=0}\Bigl[\frac{(-g_{\mu\nu})B^{\mu\nu}}{2}\Bigr],\end{eqnarray}
where $\mu$ and $\nu$ are Lorentz indices, the factor 1/2 is the average over gluon helicities, and $d^{\mu'\nu'}(q)$ is given by

\[d^{\mu'\nu'}(q)=-g^{\mu'\nu'}+(n^{\mu'}q^{\nu'}+q^{\mu'}n^{\nu'})/(nq).\]
 Here $n$ is a light-like vector that specifies the light-cone gauge. Similarly, inserting the projection operator $P^{(Q)}(q)$ for quark internal lines into Eq. (2$\cdot$2), we have

\begin{eqnarray}P^{(Q)}(q)C={\mathrm{Tr}}\Bigl[A_{\alpha\alpha'}{\Slash{q}}^{\alpha\alpha'}\Bigr]_{q^2=0}\frac{1}{4{qn}}{\mathrm{Tr}}\Bigl[{\Slash{n}}^{\beta\beta'}B_{\beta\beta'}\Bigr],\end{eqnarray}
where $\alpha, \beta, \alpha'$ and $\beta'$ are the spinor indices. Hence, using Eqs. (2$\cdot$3) and (2$\cdot$4), the projection operators are given by 
 
\begin{eqnarray}P^{(Q)}(q)=\frac{\Slash{n}}{4qn}\end{eqnarray}
and 

\begin{eqnarray}P^{(G)}(q)=-\frac{g_{\mu\nu}}{2}.\end{eqnarray}
 The second factors on the right-hand sides of Eqs. (2$\cdot$3) and (2$\cdot$4) include the collinear singularities.
 
 The diagrams contributing to the corrections at order $\alpha_s^2$ in the light-cone gauge are displayed in Fig. 1.
 
 In the jet calculus,\cite{rf:5} the minimum mass scale $M_0^2$ is used as the resolution that distinguishes one jet from two jets in the connected line for the ladder diagram. For instance, $M_0^2$ can be applied to the lines labeled ``j" and ``k" in Fig. 1. The three-body decay function arises from the contribution above $M_0^2$. Also, the contribution below $M_0^2$ is absorbed into the two-body decay.\cite{rf:5}

\begin{figure}
\epsfxsize=10cm
\centerline{\epsfbox{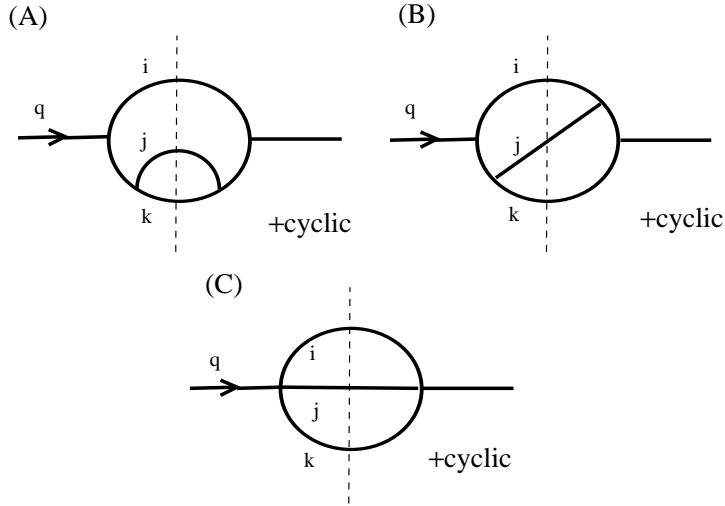}}
\caption{The diagrams contributing to the corrections at $\alpha_s^2$ order in the light-cone gauge. The labels i, j and k represent the momenta of the partons. (A) Type [L]: Ladder diagrams with two identical time-like propagators. Here, $L^{(1)}{\propto}1/t_{23}^2, L^{(2)}{\propto}1/t_{13}^2, L^{(3)}{\propto}1/t_{12}^2$. (B) Type [I]: Interference diagrams with two different time-like propagators. Here, $I^{(1)}{\propto}1/t_{12}t_{23}, I^{(2)}{\propto}1/t_{12}t_{13}, I^{(3)}{\propto}1/t_{13}t_{23}.$ (C) Four-gluon interaction for pure gluon decays.}
\label{Fig:1}
\end{figure}
 
 In this paper, we calculate the three-body decay functions for the following processes:

 Process(G1)\[g(q){\to}g(l_1)+g(l_2)+g(l_3),\]

 Process(Q1)\[q(q){\to}g(l_1)+g(l_2)+q(l_3),\]

 Process(G2)\[g(q){\to}q(l_1)+g(l_2)+\bar{q}(l_3),\]

 Process(Q2)\[q(q){\to}q(l_1)+q(l_2)+\bar{q}(l_3).\]
Here, the momentum of the mother parton with time-like virtuality is denoted by $q$. The momenta of the three daughter partons on the mass shell are denoted by $l_1,l_2$ and $l_3$. The kinematics for the variables are determined as $q^2=t$ and $l_1^2=l_2^2=l_3^2$=0. 
 
  Here, the invariants are defined as

\begin{eqnarray}
 t_{ij}=(l_i+l_j)^2, \;\;(i{\neq}j)\\
 t_{12}+t_{13}+t_{23}=t.
\end{eqnarray}
The collinear contribution of the branching vertex at $\alpha_s^2$ order extracted by the projection operator is defined as  

\begin{eqnarray}V=g^4{\int}{d\Gamma}{\P}M\frac{1}{t^2},\end{eqnarray}
where $g$ is the QCD coupling constant, $M$ is the squared matrix element, and ${\P}$ [where i=$Q$ (quark) or $G$ (gluon)] is the projection operator that is given by Eqs. (2$\cdot$5) and (2$\cdot$6). Also here, d$\Gamma$ is the three-body phase space volume element, which is given by

\begin{eqnarray}d\Gamma=(2\pi)^4\delta^{(4)}(q-l_1-l_2-l_3)\frac{1}{(2\pi)^3}\frac{d^3l_1}{2l_{10}}
\frac{1}{(2\pi)^3}\frac{d^3l_2}{2l_{20}}\frac{1}{(2\pi)^3}\frac{d^3l_3}{2l_{30}}.\end{eqnarray}
The Sudakov variable for the final state parton is defined as 

\begin{eqnarray}l_i={\alpha}_in+{\beta}_i\bar{n}+l_{iT},\;\; (n^2=\bar{n}^2=nl_{iT}=\bar{n}l_{iT}=0 )\end{eqnarray}
where $\bar{n}$ is a vector, $l_{iT}$ is the transverse momentum, and
 
\begin{eqnarray}\alpha_i=\frac{\vec{l}\,^2_{iT}}{2\beta_i\bar{n}n}.
\end{eqnarray}
The Sudakov variable for the parton with virtuality $t$ is also given by 

\begin{eqnarray}q={\alpha_q}n+\beta_q\bar{n}+q_T, 
\end{eqnarray}
where $q_T$ is the transverse momentum and we set $q_T=0,\alpha_q=\beta_q=1$.
 
 Using Eqs. (2$\cdot$11) and (2$\cdot$13), the momentum fraction $z_i$ is defined by

\begin{eqnarray}z_i=\frac{l_in}{qn}=\frac{l_in}{n\bar{n}}=\beta_i.
\end{eqnarray}
From the Sudakov variables and the momentum fractions, the phase space volume element can be written 

\begin{eqnarray}d\Gamma=\frac{1}{(2\pi)^6}\delta(q-l_i-l_j-l_k)\frac{dz_i}{2z_i}\frac{dz_j}{2z_j}d^2{\vec{l}}_{iT}d^2{\vec{l}}_{jT}d^4l_k\delta(l_k^2),\end{eqnarray} 
with $(i,j,k)=(1,2,3)$ or (1,3,2). This phase space volume element can be written in terms of d$\tilde{\Gamma}$, which is defined as 

\begin{eqnarray}d\tilde{\Gamma}=d^2\vec{l}_{iT}d^2\vec{l}_{jT}\delta(l_k^2),\end{eqnarray}
with $(i,j,k)=(1,2,3)$ or (1,3,2). Employing Eq. (2$\cdot$16), we find that d$\Gamma$ takes the form
 
\begin{eqnarray}d\Gamma=\frac{1}{(8\pi^2)^2}\frac{1}{4\pi^2X}\delta(1-z_1-z_2-z_3)dz_1dz_2dz_3d\tilde{\Gamma},\end{eqnarray} 
 where $X=z_iz_j$ and $(i,j)=(1,2)$ or (1,3).

 From Eq. (2$\cdot$17), the branching vertex function is given by

\begin{eqnarray}V=\Bigl(\frac{\alpha_s}{2\pi}\Bigr)^2\delta(1-z_1-z_2-z_3)dz_1dz_2dz_3J\frac{dt}{t},\end{eqnarray}
where $\alpha_s=g^2/4\pi$, and 
 
\begin{eqnarray}J=\frac{1}{4\pi^2X}d\tilde{\Gamma}{\P}M\frac{t}{t^2}.\end{eqnarray}   
The quantity $J$ is called the ``decay function." It consists of a single logarithmic term and a double logarithmic term, due to the collinear singularity. The contribution of the double logarithmic term should be subtracted from that of the three-body decay function to avoid  double counting.\cite{rf:4,rf:5,rf:10} In order to determine the distribution for $\delta(1-z_1-z_2-z_3)dz_1dz_2dz_3$ in Eq. (2$\cdot$18), we integrate over the phase space with volume element $d\tilde{\Gamma}$.
 
 We now briefly explain the calculation of the squared matrix element ${\P}M$. We used REDUCE \footnote{REDUCE was developed by A.~Hearn.} to calculate this squared matrix element. Replacing the inner product of the momenta in ${\P}M$ by the momentum fractions and the invariants, ${\P}M$ can be calculated. [See Eqs. (A$\cdot$6) and (A$\cdot$23) in Appendix A.]
 
Integrating ${\P}M$ over the azimuthal angle, it can be verified that there are no terms proportional to $t^2$ in the numerator of ${\P}M$.
 [See Eq. (A$\cdot$11) in Appendix A.] The ladder diagrams cannot have terms that are more singular than logarithmic order. Furthermore, the condition that the interference diagram has no mass singularity was confirmed. We examined whether or not the terms proportional to the invariants in the numerator of ${\P}M$ satisfy the condition of having no mass singularity. [See Eq. (A$\cdot$32) in Appendix A.] The details of the calculations are presented in Appendix A. 
 
\section{Calculated results}
 Here the calculated results for the decay functions in the light-cone gauge are presented. The ladder diagrams include the terms $L_L^{(k)}$, which are the convolutions of the Altarelli-Parisi splitting functions.\cite{rf:11} Each type of calculated decay function is represented as 
 
\begin{eqnarray}J^{[L^{(k)}]}&=& \int^{t}_{{M_0}^2}L_L^{(k)}\frac{dt_{ij}}{t_{ij}}+L_L^{(k)}\log{y_k}+L_N^{(k)},\\J^{[I^{(k)}]}&=& I^{(k)}_L{\log}
\frac{y_iy_j}{z_k}+I^{(k)}_N, \;\;(i,j,k=1,2,3,i{\neq}j{\neq}k)\end{eqnarray} 
where $y_i=1-z_i\;\;(i=1,2,3)$ and $O(M_0^2/t)$ terms are ignored. $M_0$ is the minimum mass scale of the phase space integration.

 Integrating over $t_{ij}$ for Type [L] and summing over all contributions from $J^{[L^{(1)}]}$ to $J^{[I^{(3)}]}$, we obtain\cite{rf:7}
\begin{eqnarray}\sum^{I^{(3)}}_{i=L^{(1)}}J^{[i]}=V_{LL}\log\frac{t}{M_0^2}+V_{NLL},\end{eqnarray}
 where
\begin{eqnarray}V_{LL}=L_L^{(1)}+L_L^{(2)}+L_L^{(3)}\end{eqnarray}
 and
\begin{eqnarray}V_{NLL}=V_L+V_N,\end{eqnarray}
 with
\begin{eqnarray}V_L&=& \sum^3_{i=1}L_L^{(i)}{\log}y_i+\sum^3_{j,k,l=1}I_L^{(j)}{\log}\frac{y_ky_l}{z_j}, \;\;(j{\neq}k{\neq}l)\end{eqnarray}
\begin{eqnarray}V_N=\sum^3_{i=1}L_N^{(i)}+\sum^3_{j=1}I_N^{(j)}.\end{eqnarray}
The terms $V_{LL}$ and $V_{NLL}$ in Eq. (3$\cdot$3) are the contributions to LL order and NLL order at order $\alpha_s^2$, respectively. The term $V_{NLL}$, called the three-body decay function, is constructed from the logarithmic term $V_L$ and the non-logarithmic term $V_N$. 

 The three-body decay functions are calculated in Refs.~\citen{rf:4} and \citen{rf:5}. We compare the results given there with the new results reported in this paper. The results for Processes (G1) and (Q2) in this paper are consistent with the calculated results in Refs.~\citen{rf:4} and \citen{rf:5}. While the results for Process (Q1) obtained in Refs.~\citen{rf:4} and \citen{rf:5} are quite different, our result agrees with that in Ref.~\citen{rf:4}. The results, excluding the logarithmic terms for $I^{(1)}$ and $I^{(3)}$, for Process (G2) obtained in this paper are consistent with those in both Refs.~\citen{rf:4} and \citen{rf:5}.
 
 Using the notation defined in $\S$2 and 3, the results for the logarithmic terms of the interference diagrams $I^{(1)}$ and $I^{(3)}$ for Process (G2) are presented. The results used in NLLJET\cite{rf:4} are 

\begin{eqnarray}& &I_L^{(1)}\log\frac{y_2y_3}{z_1}+I_L^{(3)}\log\frac{y_1y_2}{z_3}\nonumber  \\&=& -C_AT_R\Bigl[\frac{z_1^2+z_3^2}{z_2}-\frac{z_1^2+z_3^2+1}{y_2}+1-z_1+z_3\Bigr]\log\frac{y_2y_3}{z_3}\nonumber \\& &{}-C_AT_R\Bigl[\frac{z_1^2+z_3^2}{z_2}-\frac{z_1^2+z_3^2+1}{y_2}+1+z_1-z_3\Bigr]\log\frac{y_1y_2}{z_1}.\end{eqnarray}
For comparison, the results given in Ref.~\citen{rf:5} are 
\begin{eqnarray}& &I_L^{(1)}\log\frac{y_2y_3}{z_1}+I_L^{(3)}\log\frac{y_1y_2}{z_3}\nonumber  \\&=& -C_AT_R\Bigl[\frac{z_1^2+z_3^2}{z_2}-\frac{z_1^2+z_3^2+1}{y_2}+1+z_1-z_3\Bigr]\log\frac{y_2y_3}{z_1}\nonumber \\& &{}-C_AT_R\Bigl[\frac{z_1^2+z_3^2}{z_2}-\frac{z_1^2+z_3^2+1}{y_2}+1-z_1+z_3\Bigr]\log\frac{y_1y_2}{z_3}.\end{eqnarray}
The new results obtained in this paper are the following:
\footnote{As discussed in Ref.~\citen{rf:7}, using the results given in Ref.~\citen{rf:8}, only Eq. (3$\cdot$10) satisfies the relation of the crossing symmetry between the time-like branching process and the space-like branching process.},\cite{rf:8}

\begin{eqnarray}& &I_L^{(1)}\log\frac{y_2y_3}{z_1}+I_L^{(3)}\log\frac{y_1y_2}{z_3}\nonumber  \\&=& -C_AT_R\Bigl[\frac{z_1^2+z_3^2}{z_2}-\frac{z_1^2+z_3^2+1}{y_2}+1-\frac{z_1-z_3}{y_2}\Bigr]\log\frac{y_2y_3}{z_1}\nonumber \\& &{}-C_AT_R\Bigl[\frac{z_1^2+z_3^2}{z_2}-\frac{z_1^2+z_3^2+1}{y_2}+1+\frac{z_1-z_3}{y_2}\Bigr]\log\frac{y_1y_2}{z_3}.\end{eqnarray}

\begin{figure}
\epsfxsize=10cm
\centerline{\epsfbox{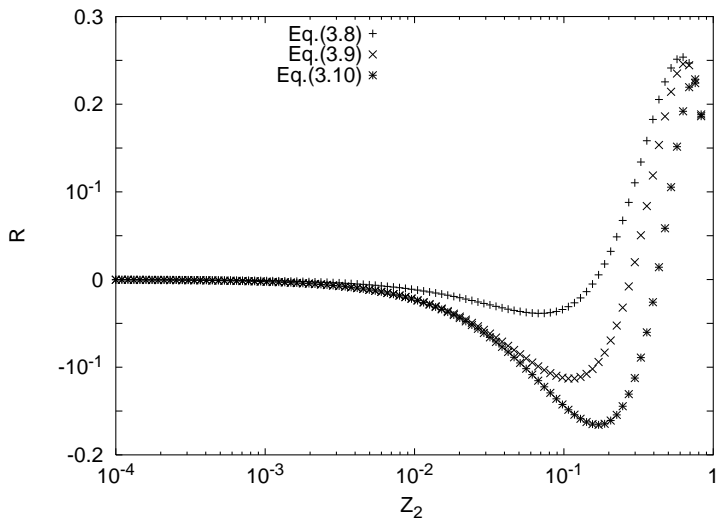}}
\caption{The $z_2$ dependence of $R$ described by Eq. (3$\cdot11$) at $z_3=10^{-1}$ for Process (G2). The symbols correspond to the different numerators for $R$ corresponding to Eqs. (3$\cdot$8)--(3$\cdot$10).}
\label{Fig:2}
\end{figure} 

 We next examine the numerical results for the calculated three-body decay functions given in Refs.~\citen{rf:4} and \citen{rf:5} and the new results of this paper. We define the ratio $R$ as 
 
 \begin{eqnarray}R=\frac{I_L^{(1)}\log\frac{y_2y_3}{z_1}+I_L^{(3)}\log\frac{y_1y_2}{z_3}}{V_{LL}}.\end{eqnarray}
Equations (3$\cdot$8)--(3$\cdot$10) are inserted into the numerator in Eq. (3$\cdot11$). As shown in Fig. 2, the numerical result for Eq. (3$\cdot$10) differs significantly from the results given in Refs.~\citen{rf:4} and \citen{rf:5} in the region [$z_2{\geq}10^{-2}$] for $z_3=10^{-1}$.

\section{Numerical results}
 In order to investigate the properties of the three-body decay functions in the soft regions, some numerical quantities are examined. These properties are useful to construct and improve the parton shower model. 
 
 We define the following ratios:
 
\begin{eqnarray}R_1=\frac{V_{NLL}}{V_{LL}},\end{eqnarray}
\begin{eqnarray}R_2=\frac{V_L}{V_{LL}},\end{eqnarray}
\begin{eqnarray}R_3=\frac{V_N}{V_L}.\end{eqnarray}
 
 If the absolute value of $R_1$ is large, the contribution of the NLL order becomes important, because the contribution of the term $V_{NLL}$ is large relative to that of the term $V_{LL}$ at $\alpha_s^2$ order. If $R_1{\simeq}R_2$, the dominant contribution of $V_{NLL}$ is the logarithmic term $V_L$. The ratio of the contribution of the logarithmic term $V_L$ to that of the non-logarithmic term $V_N$ is represented by $R_3$.

\subsection{$z_2$ dependences of $R_1$ and $R_2$}
 In this subsection, we focus on the relation between the contributions of the logarithmic term $V_L$ and the non-logarithmic term $V_N$. In Figs. 3--5,  the gluon momentum fraction $z_2$ dependences of $R_1$ and $R_2$ are presented at $z_3=0.5, 10^{-1}, 10^{-2}, 10^{-3}$. Here, $R_1$ and $R_2$ are depicted by the solid curves and the crosses, respectively.

 First, we examine the $z_2$ dependences of $R_1$ and $R_2$ for Process (G1), in which a gluon decays into three gluons. In Fig. 3, $R_1{\simeq}R_2$ in most of region depicted. The absolute values of $R_1$ and $R_2$ in the double soft gluon region [$z_2{\simeq}z_3{\simeq}y_1/2{\ll}1$] are small relative to those in the other region. 
 
 The numerical result for Process (Q1), in which a quark decays into one quark and two gluons, is shown in Fig. 4. Here, $R_1{\simeq}R_2$, except in the region of the small quark momentum fraction $z_3$ and the hard gluon momentum fraction $z_2$ (soft quark region). The contribution of the non-logarithmic term $V_N$ is important in the soft quark region. As the quark momentum fraction $z_3$ is hard and the gluon momentum fraction $z_2$ is small (soft gluon region), the absolute values of $R_1$ and $R_2$ are large. In addition, the absolute values of $R_1$ and $R_2$ in the soft quark region are small. 
 
\begin{figure}
\epsfxsize=10cm
\centerline{\epsfbox{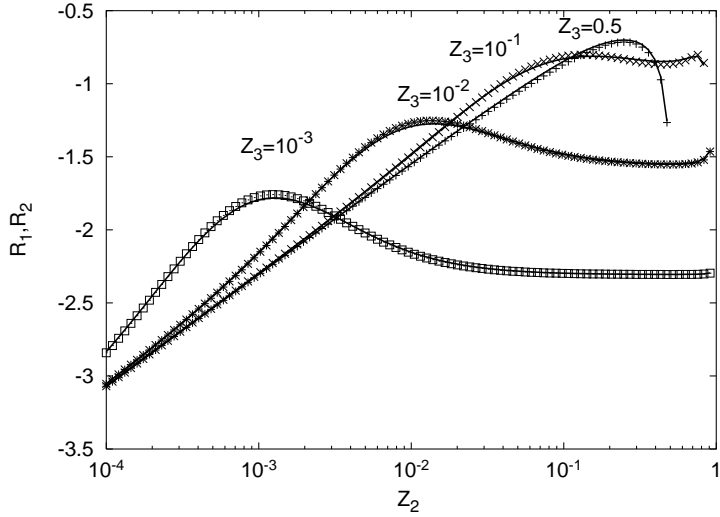}}
\caption{The $z_2$ dependences of the ratios $R_1$ and $R_2$ for Process (G1). The solid curves and the crosses denote $R_1$ and $R_2$. The symbols $\Box$, $\ast$, $\times$ and + correspond to the values $10^{-3}$, $10^{-2}$, $10^{-1}$, 0.5 for $z_3$. }
\label{Fig:3}
\end{figure}

 The numerical result for Process (G2), in which a gluon decays into one quark, one anti-quark and one gluon, is presented in Fig. 5. In Process (G2), $R_1{\simeq}R_2$ in the strong ordering region [$z_2{\ll}z_3{\ll}z_1$]. As the anti-quark momentum fraction $z_3$ is soft and the gluon momentum fraction $z_2$ is hard (soft anti-quark region), the absolute value of $R_1$ is large relative to that of $R_2$. Thus, the contribution of the non-logarithmic term gives an important contribution to the soft anti-quark region.

 In the following subsections, we examine the properties of the three-body decay functions in each of the $R_1{\simeq}R_2$ regions and the soft fermion (quark or anti-quark) regions.

\begin{figure}
\epsfxsize=10cm
\centerline{\epsfbox{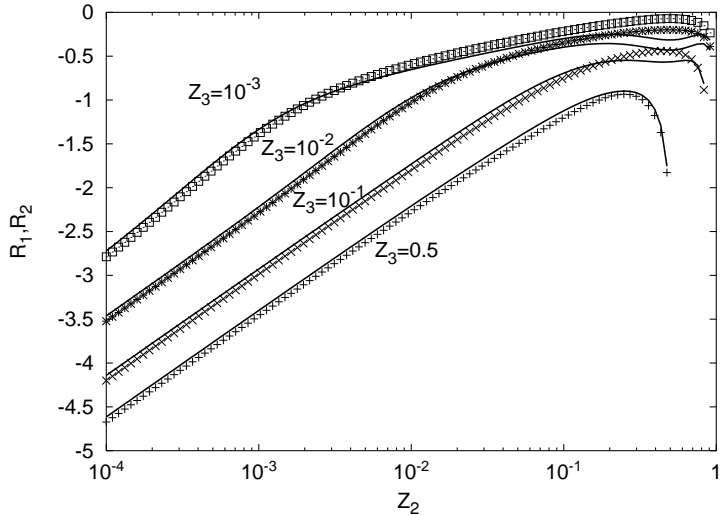}}
\caption{The $z_2$ dependences of the ratios $R_1$ and $R_2$ for Process (Q1). The solid curves and the crosses denote $R_1$ and $R_2$. The symbols $\Box$, $\ast$, $\times$ and + correspond to the values $10^{-3}$, $10^{-2}$, $10^{-1}$, 0.5 for $z_3$.}
\label{Fig:4}
\end{figure}
\begin{figure}
\epsfxsize=10cm
\centerline{\epsfbox{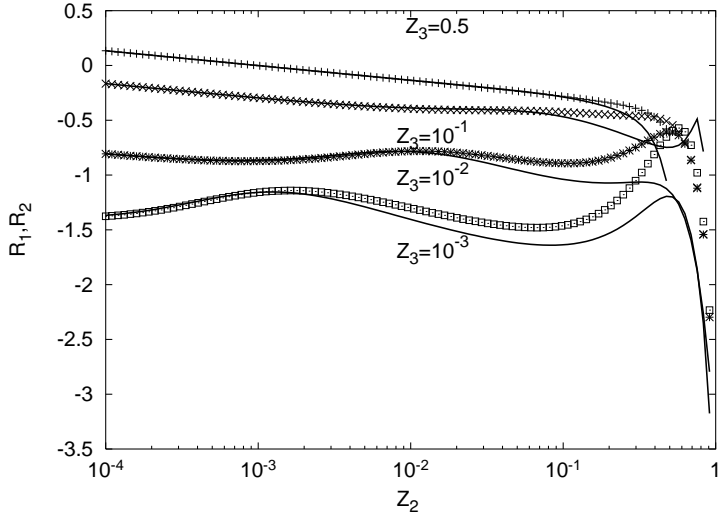}}
\caption{The $z_2$ dependences of the ratios $R_1$ and $R_2$ for Process (G2). The solid curves and the crosses denote $R_1$ and $R_2$. The symbols $\Box$, $\ast$, $\times$ and + correspond to the values $10^{-3}$, $10^{-2}$, $10^{-1}$, 0.5 for $z_3$.}
\label{Fig:5}
\end{figure}

\subsection{Cancellation of the non-logarithmic term}
 As in Ref. ~\citen{rf:7}, we examine the cancellations of the non-logarithmic terms $V_N$ in detail. Figures 6--8 confirm the structure of the cancellations of the non-logarithmic terms $V_N$ in the region in which $R_1{\simeq}R_2$. For each process, the structure of the cancellation of the non-logarithmic term $V_N$ is different. For these processes we have the following relations:
 
 Process (G1)
 
 \begin{eqnarray}\bigl(I_N^{(1)}+I_N^{(2)}+I_N^{(3)}\bigr){\simeq}\bigl(L_N^{(1)}+L_N^{(2)}+L_N^{(3)}\bigr),\end{eqnarray}
 
 Process (Q1)
 
 \begin{eqnarray}\bigl(I_N^{(1)}+I_N^{(2)}\bigr){\simeq}\bigl(L_N^{(1)}+L_N^{(2)}+L_N^{(3)}+I_N^{(3)}\bigr),\end{eqnarray}
 
 Process (G2)
 
 \begin{eqnarray}\bigl(I_N^{(1)}+I_N^{(3)}\bigr){\simeq}\bigl(L_N^{(1)}+L_N^{(2)}+L_N^{(3)}+I_N^{(2)}\bigr).\end{eqnarray}
 
 In the case of Process (G1), Eq. (4$\cdot$4) implies that the contributions for the non-logarithmic terms of the interference diagrams are canceled by those of the ladder diagrams. Equations (4$\cdot$5) and (4$\cdot$6) imply that the contributions of the three ladder diagrams and one interference diagram for Processes (Q1) and (G2) cancel those of the non-logarithmic terms for the interference diagrams. 
 
\begin{figure}
\epsfxsize=10cm
\centerline{\epsfbox{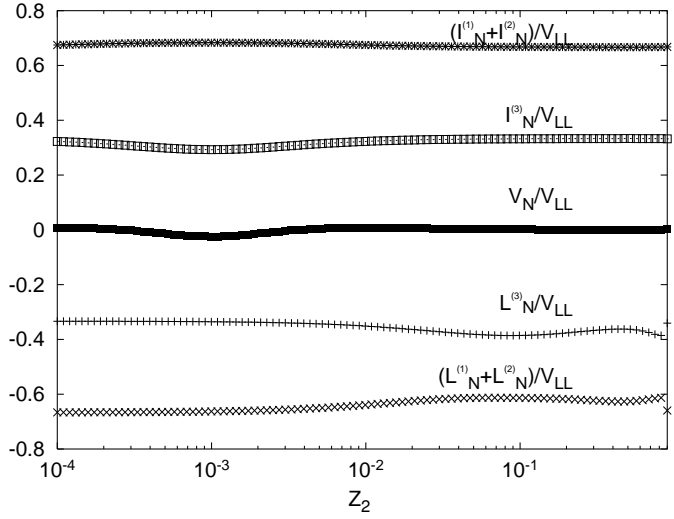}}
\caption{The $z_2$ dependences of non-logarithmic terms for Process (G1) at $z_3=10^{-1}$. The notation ($L_N$, etc.) is defined by Eq. (3$\cdot$7) in the text. Here, $z_2$=$10^{-4}$--0.9.}
\label{Fig:6}
\end{figure} 
\begin{figure}
\epsfxsize=10cm
\centerline{\epsfbox{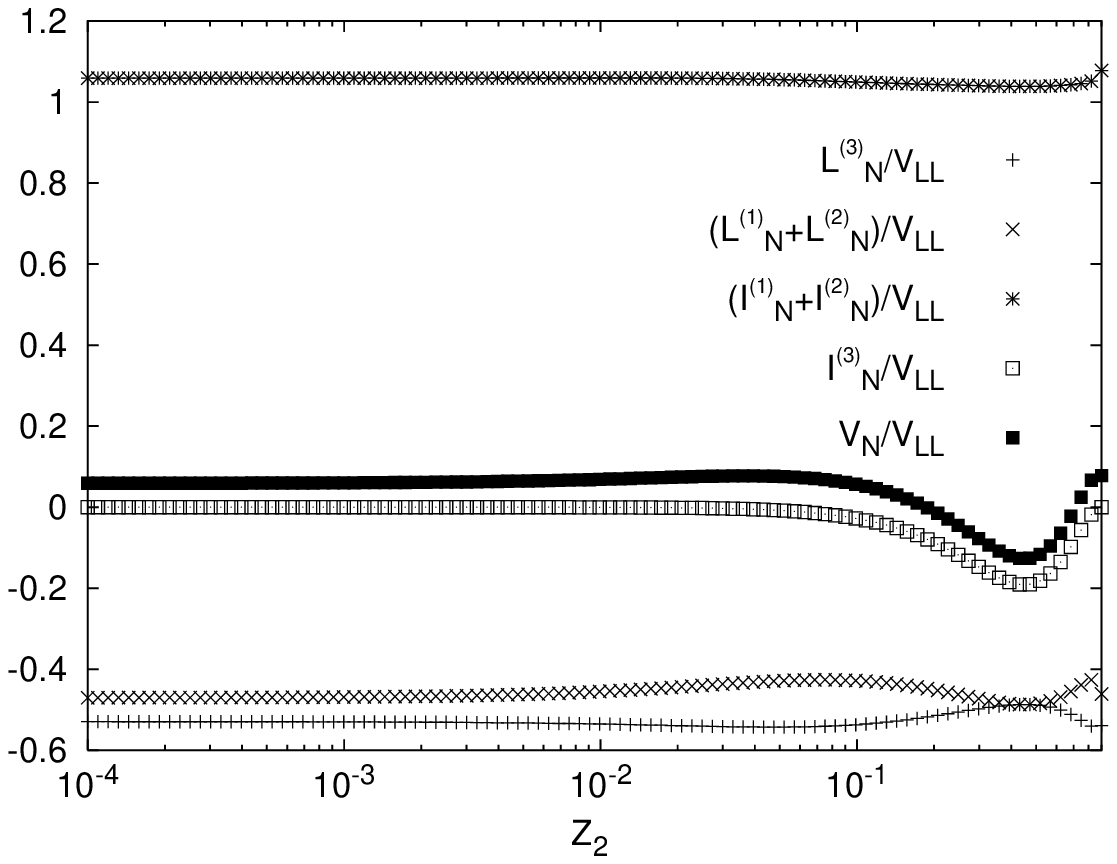}}
\caption{The $z_2$ dependences of non-logarithmic terms for Process (Q1) at $z_3=10^{-1}$. The notation ($L_N$, etc.) is defined by
Eq. (3$\cdot$7) in the text. Here, $z_2$=$10^{-4}$--0.9.}
\label{Fig:7}
\end{figure}
\begin{figure}
\epsfxsize=10cm
\centerline{\epsfbox{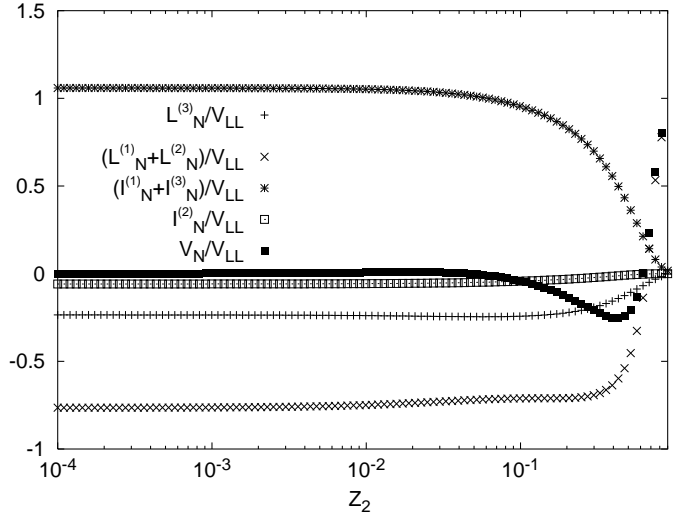}}
\caption{The $z_2$ dependences of non-logarithmic terms for Process (G2) at $z_3=10^{-1}$. The notation ($L_N$, etc.) is defined by Eq. (3$\cdot$7) in the text. Here,$z_2$=$10^{-4}$-0.9.}
\label{Fig:8}
\end{figure}

\subsection{Soft fermion $[$quark--anti-quark$]$ region}
 As mentioned above, we showed that the dominant contribution of $V_{NLL}$ is the logarithmic term $V_L$, except in the soft quark region for Process (Q1) and the soft anti-quark region for Process (G2). First, we examine the soft quark limit for Process (Q1). In Fig. 9, there are some peaks of the ratio $R_3$ present in the soft quark region [$z_3{\leq}10^{-4}$]. In order to understand the meaning of the peaks in Fig. 9, we examine the structures of the logarithmic terms $V_L$ in the soft quark region in detail.

\begin{figure}
\epsfxsize=10cm
\centerline{\epsfbox{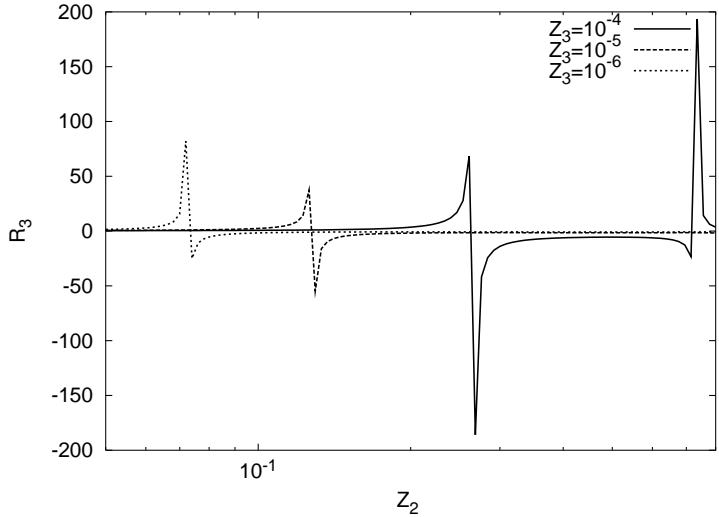}}
\caption{The $z_2$ dependence of the ratio $R_3$ with $z_3=10^{-4},10^{-5}$ and $10^{-6}$ for Process (Q1). Here, $z_2$=0.05--0.8.}
\label{Fig:9}
\end{figure}
 
\begin{figure}
\epsfxsize=10cm
\centerline{\epsfbox{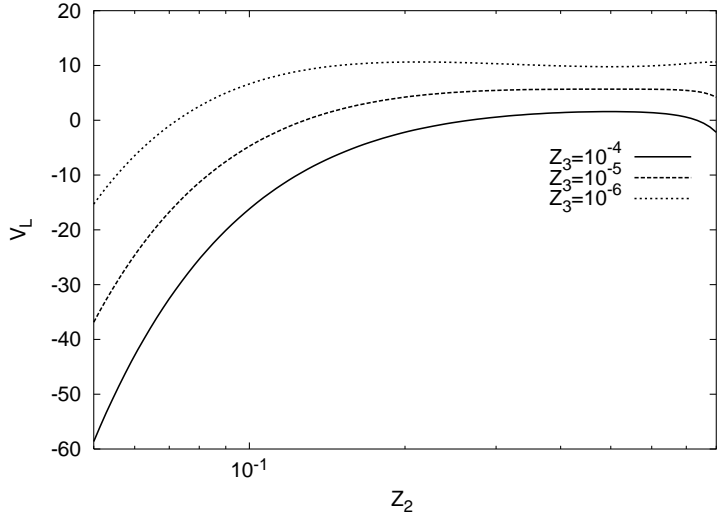}}
\caption{The $z_2$ dependence of $V_L$ for Process (Q1) with $z_3=10^{-4},10^{-5}$ and $10^{-6}$. Here, $z_2$=0.05--0.8.}
\label{Fig:10}
\end{figure}

 In Fig. 10, the contribution of the logarithmic term $V_L$ for the soft quark region crosses zero at some points. As the quark momentum fraction $z_3$ is small, the contribution of the logarithmic term crosses zero in the small $z_2$ region. 
 
 Next, we show how the logarithmic term $V_L$ crosses zero. The momentum fractions in the soft quark region are given by
 
 \begin{eqnarray}y_3{\simeq}1,\hspace{5mm}y_1{\simeq}z_2 ,\hspace{5mm}y_2{\simeq}z_1.\end{eqnarray}
 From Eq. (4$\cdot$7), the logarithmic term $V_L$ can be approximated as
  
 \begin{eqnarray}V_L{\simeq}L_L^{(1)}{\log}z_2+L_L^{(2)}{\log}z_1+I_L^{(3)}{\log}\frac{z_1z_2}{z_3},\end{eqnarray}
with
 
 \begin{eqnarray}L_L^{(1)}{\simeq}C_F^2\frac{1+z_2^2}{z_2y_2},L_L^{(2)}{\simeq}C_F^2\frac{1+y_2^2}{z_2y_2}\end{eqnarray}
and

\begin{eqnarray}I_L^{(3)}{\simeq}-2C_F\bigl(C_F-\frac{1}{2}C_A\bigr)\frac{1}{z_2y_2}.\end{eqnarray}
Using Eqs. (4$\cdot$9) and (4$\cdot$10), the dominant singular term in Eq. (4$\cdot$8) can be given as

\begin{eqnarray}V_L{\simeq}\frac{C_F^2}{z_2y_2}\log\frac{z_3^2}{z_1z_2}+\frac{C_FC_A}{z_2y_2}\log\frac{z_1z_2}{z_3}.\end{eqnarray} 

\begin{figure}
\epsfxsize=10cm
\centerline{\epsfbox{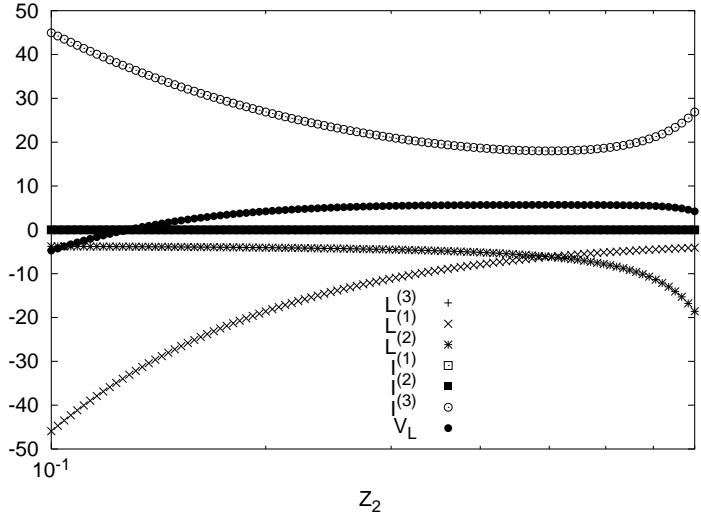}}
\caption{The $z_2$ dependences of the logarithmic terms of the ladder diagrams $L^{(k)}$ and the interference diagrams $I^{(k)}$ for Process (Q1) with $z_3=10^{-5}$. The logarithmic terms for the ladder diagrams $L^{(k)}$, etc. are given by Eq. (3$\cdot$6). Here, $z_2$=$10^{-1}$--0.8.}
\label{Fig:11}
\end{figure}

In Fig. 11, since the contributions for these two terms cancel, the logarithmic term $V_L$ crosses zero. The first term and the second term in Eq. (4$\cdot$11) arise from the ladder diagrams and the interference diagram, respectively. 

As seen from Fig. 12, the peaks of the ratios $R_3$ for Process (G2) appear near $z_2=0.6$. As for Process (Q1), it is expected that the contributions of the logarithmic terms are small in this region.

\begin{figure}
\epsfxsize=10cm
\centerline{\epsfbox{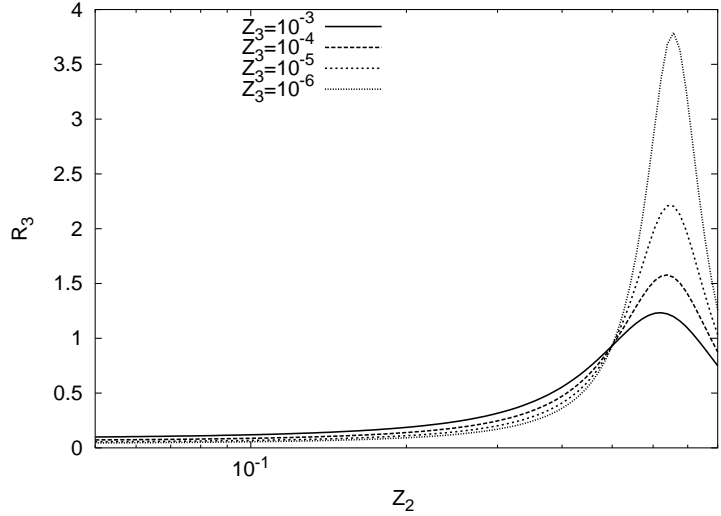}}
\caption{The $z_2$ dependence of the ratio $R_3$ for Process (G2) with $z_3=10^{-3},10^{-4},10^{-5}$ and $10^{-6}$. Here, $z_2$=0.05--0.8.}
\label{Fig:12}
\end{figure}
 
 In Fig. 13, because the sign of the numerical value of the logarithmic term for the ladder diagram $L^{(2)}$ of Process (G2) is opposite to that of the sum of the logarithmic terms for the interference diagrams $I^{(1)}$ and $I^{(3)}$, the contribution of the logarithmic term $V_L$ is small. It is easy to derive this structure analytically. The logarithmic term $V_L$ in Eq. (4$\cdot$8), taking account of the limit $z_2{\to}1$ and the relation $z_3{\ll}z_2$, is given by
 
\begin{eqnarray}V_L{\simeq}L_L^{(2)}{\log}y_2+I_L^{(3)}\log\frac{y_2}{z_3},\end{eqnarray}
with

 \begin{eqnarray}L_L^{(2)}{\simeq}\frac{2C_AT_R}{y_2^2}\hspace{5mm}I_L^{(3)}{\simeq}C_AT_R\Bigl[\frac{2z_2-1}{z_2y_2}\Bigr].\end{eqnarray}
\begin{figure}
\epsfxsize=10cm
\centerline{\epsfbox{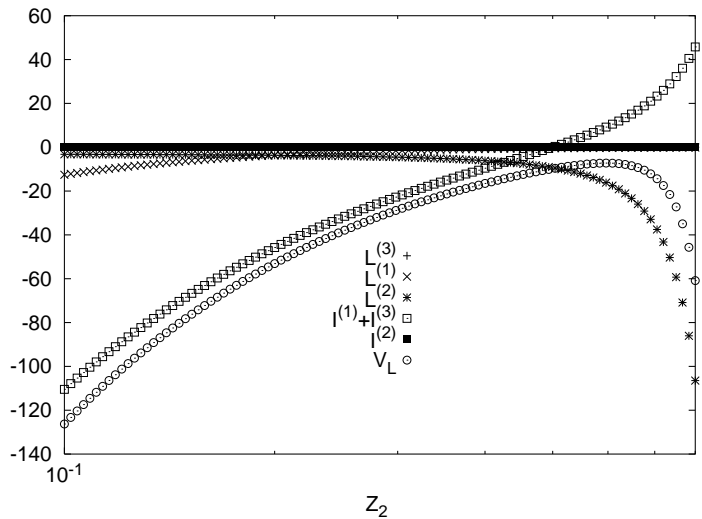}}
\caption{The $z_2$ dependences of the logarithmic terms for the ladder diagrams $L^{(k)}$ and the interference diagrams $I^{(k)}$ for Process (G2) with $z_3=10^{-5}$. The logarithmic terms for $L^{(k)}$, etc, are given by Eq. (3$\cdot$6). Here, $z_2$=$10^{-1}$--0.8.}
\label{Fig:13}
\end{figure}
In Eq. (4$\cdot$13), the contribution of the logarithmic term for the interference diagram $I^{(3)}$ is positive for $z_2{\textgreater}0.5$, which agrees with the behavior depicted in Fig. 13.

 In the next section, using the results obtained in this section, we discuss how to treat the contributions of the three-body decay functions in the parton shower model.
 
\section{Comments on the angular ordering}
\subsection{Angular ordering for the soft gluon region}
 Before the modifications of the three-body decay functions in the parton shower model are discussed, we fit the approximated forms to the numerical results in order to check the results obtained in $\S$4. In the region of the soft gluon momentum fraction $z_2$, the momentum fractions are given by

\begin{eqnarray}y_1{\simeq}z_3,y_2{\simeq}1,y_3{\simeq}z_1.\end{eqnarray}
By use of Eq. (5$\cdot$1), the logarithmic term $V_L$ is approximated by

\begin{eqnarray}V_L{\simeq}L_L^{(1)}{\log}z_3+L_L^{(3)}{\log}z_1+I_L^{(2)}\log\frac{z_1z_3}{z_2},\end{eqnarray}
with

\begin{eqnarray} L_L^{(1)}{\simeq}L_L^{(3)}{\simeq}-I_L^{(2)}{\simeq}4C_A^2K(z_1)\frac{1}{z_2}\hspace{5mm}\end{eqnarray}
for Process (G1), 
 
\begin{eqnarray}L_L^{(1)}{\simeq}2C_F^2P(y_1)\frac{1}{z_2},L_L^{(3)}{\simeq}-I_L^{(2)}{\simeq}2C_FC_AP(y_1)\frac{1}{z_2}\end{eqnarray}
for Process (Q1), and

\begin{eqnarray}L_L^{(1)}{\simeq}L_L^{(3)}{\simeq}2C_FT_RH(z_1)\frac{1}{z_2},-I_L^{(2)}{\simeq}4T_R\bigl(C_F-\frac{1}{2}C_A\bigr)H(z_1)\frac{1}{z_2}\end{eqnarray}
for Process (G2), where $P(z)=(1+z^2)/(1-z),K(z)=1/z+1/(1-z)-2+z(1-z)$ and $H(z)=z^2+(1-z)^2$ are Altarelli-Parisi splitting functions.\cite{rf:11} The contributions to Eqs.~(5$\cdot$3)--(5$\cdot$5) agree with the solid curves $R_1$ in Figs.~2--4. Also, since the contributions of the non-logarithmic terms $V_N$ in the one soft gluon regions are small relative to those of the logarithmic terms $V_L$ (Figs. 2--4), we ignore the contributions of the non-logarithmic terms $V_N$. 
 
 Equations (5$\cdot$1)--(5$\cdot$4) offer a physical interpretation of the contributions of the three-body decay functions for the parton shower model. The singular factor $\log{z}/z$ of the three-body decay function in the one soft gluon limit comes from the contribution of the interference diagram. This contribution due to the interference diagram causes the perturbative expansion to break down. In addition, as shown in Figs. 3 and 4, the contributions of the three-body decay functions are negative for small $z_2$ for Processes (G1) and (Q1). Since the parton shower model generates momentum fractions of the branching partons by decay functions, the contributions of the decay functions are required to be positive. In order to avoid these difficulties, it is necessary for the singular term to be absorbed into the kinematical boundary of the phase space for the two-body decay by the angular ordering. \cite{rf:4,rf:9} 

 We expect that many soft gluons will be produced in the TeV energy region. There should exist the situation that the momentum fractions of the two gluons are soft. For this reason, we examine the relation between the double soft gluon limit and the angular ordering reported in Ref.~\citen{rf:4}. We obtain momentum fractions for the double soft gluon limit that satisfy the following:

\begin{eqnarray}z_1{\simeq}z_2{\simeq}\frac{y_3}{2},\hspace{5mm}z_3{\simeq}1.\end{eqnarray}
 As shown in Figs. 3 and 4, $R_1{\simeq}R_2$ in the double soft gluon region. The dominant contribution of $V_{NLL}$ in this region is the logarithmic term $V_L$. It yields
 
\begin{eqnarray}V_L&{\simeq}&L_L^{(3)}{\log}y_3.\end{eqnarray}
Therefore, we calculate the $L_L^{(3)}$ term in the double soft gluon limit for Process (G1) and (Q1). We find

\begin{eqnarray}L_L^{(3)}{\simeq}9C_A^2\frac{1}{y_3^2}\end{eqnarray}
for Process(G1) and 

\begin{eqnarray}L_L^{(3)}{\simeq}\frac{9}{2}C_FC_AP(z_3)\frac{1}{y_3}\end{eqnarray}
for Process(Q1). The angular ordering in the double soft limit reported in Ref.~\citen{rf:4} satisfies these approximate formulas for the double soft gluon limit, Eqs. (5$\cdot$8) and (5$\cdot$9). 

\subsection{New restriction introduced by the kinematical boundary in the phase space for Process $($G2$)$} 
 As shown in Fig. 5, the contribution of the three-body decay function for Process (G2) is negative in the soft anti-quark region. For Process (G2), both the logarithmic terms and the non-logarithmic terms for the ladder diagrams were absorbed into the kinematical boundaries of the phase spaces for the two-body decays to maintain the positivity of the three-body decay function.\cite{rf:4}

 Following Ref.~\citen{rf:4}, the formulas of the non-logarithmic terms for the ladder diagrams are written 

\begin{eqnarray}L_N^{(k)}=-L_L^{(k)}+L'.\end{eqnarray}
Therefore, a part of the non-logarithmic terms can be absorbed into the kinematical boundaries of the phase spaces for the two-body decays. By use of Eq. (5$\cdot$10), we obtain

 \begin{eqnarray}& &\int^t_{M_0^2}L_L^{(k)}\bigl(1-\frac{t_{ij}}{t}\bigr)\frac{dt_{ij}}{t_{ij}}+L_L^{(k)}\log{y_k}+L'\nonumber \\&=&{} \int^{y_kt}_{M_0^2}L_L^{(k)}\bigl(1-\frac{t_{ij}}{ty_k}\bigr)\frac{dt_{ij}}{t_{ij}}+L',\end{eqnarray}
with $(i,j)=(1,2),(1,3)$ or (2,3). The contribution of $O$($M_0^2$/t) is absorbed into the two-body decay. 
 
 We corrected a part of the results for Process (G2) in $\S$3. The logarithmic terms for the interference diagrams $I^{(1)}$  and $I^{(3)}$ used in NLLJET are replaced by Eq. (3$\cdot$10). Now, we need to examine the contribution of the three-body decay function for Process (G2) given in Ref.~\citen{rf:4} modified by Eq. (5$\cdot$11). It is given by

\begin{subequations}
\begin{equation}W_{qg\bar{q}}=\frac{V_{NLL}-\displaystyle\sum^3_{i=1}\bigl[L_L^{(i)}\log{y_i}+L_N^{(i)}(CV)\bigr]}{V_{LL}},\end{equation}
with

\begin{eqnarray}\sum^3_{i=1}L_N^{(i)}(CV)&=&-C_FT_RH(y_1)P\bigl(\frac{z_3}{y_1}\bigr)\frac{1}{y_1}-C_FT_RH(y_3)P\bigl(\frac{z_1}{y_3}\bigr)\frac{1}{y_3}\nonumber \\& &-2C_AT_RK(z_2)H\bigl(\frac{z_1}{y_2}\bigr)\frac{1}{y_2},\end{eqnarray}\end{subequations}
where $L_N^{(i)}(CV)$ are the convolutions of the Altarelli-Parisi splitting functions\cite{rf:11} in the non-logarithmic terms for the ladder diagrams. We call the function in Eq. (5$\cdot$12a) the modified three-body decay function. We compare this decay function with the LL term in order to explore the contribution at the three-body level. Both the first term and the second term in Eq. (5$\cdot$12b) differ from the corresponding results in Ref.~\citen{rf:4}. The results in Ref.~\citen{rf:4} are not the convolutions of the Altarelli-Parisi splitting functions. They are given by
 
\begin{eqnarray}C_FT_R\Bigl[H(z_1)P\bigl(\frac{z_1}{y_3}\bigr)\frac{1}{y_3}+H(z_3)P\bigl(\frac{z_3}{y_1}\bigr)\frac{1}{y_1}\Bigr].\end{eqnarray}
Although the numerical difference between Eq. (5$\cdot$13) and the convolution term in Eq. (5$\cdot$12b) is small, we adopt the result Eq. (5$\cdot$12b).

 As shown in Fig. 14, substituting Eq. (3$\cdot$10) into Eq. (5$\cdot$12a), the contribution of Eq. (5$\cdot$12a) is negative for the anti-quark momentum fraction $z_3=10^{-3}$. Thus, the restriction introduced by the kinematical boundary of the phase space for the two-body decay with the modifications of the non-logarithmic terms for Process (G2) given in Ref.~\citen{rf:4} cannot be used in the region satisfying $z_3{\ll}z_2{\ll}z_1$ for the parton shower model.

 In order to cancel the negativity of the contribution from the modified three-body decay function [Eq. (5$\cdot$12a)] in the soft anti-quark region for Process (G2), we should construct another method. Since the negativity of the modified three-body decay function comes from the soft anti-quark region, we investigate the dominant term in the soft anti-quark region. In the soft anti-quark $z_3$ region, the logarithmic term $V_L$ is given by Eq. (4$\cdot$8), and the terms proportional to each of the logarithmic terms in $V_L$ can also be approximated by
\begin{eqnarray}L_L^{(1)}&{\simeq}&C_FT_RH(y_2)\frac{1}{z_2},\\L_L^{(2)}&{\simeq}&2C_AT_R\frac{1}{z_2y_2},\\I_L^{(3)}&{\simeq}&-C_AT_R\frac{1}{z_2y_2}.\end{eqnarray}
\begin{figure}
\epsfxsize=10cm
\centerline{\epsfbox{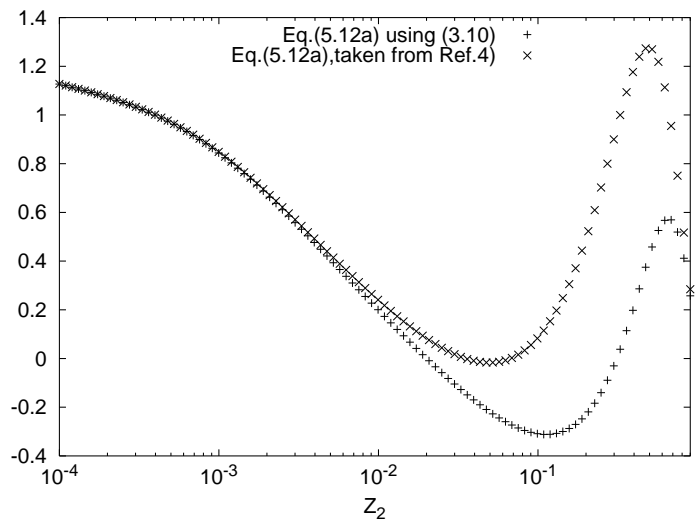}}
\caption{The $z_2$ dependences of Eq. (5$\cdot$12a), taken from Ref. 4) and Eq. (5$\cdot$12a) using Eq. (3$\cdot$10) for Process (G2) with $z_3=10^{-3}$. Here, $z_2$=$10^{-4}$--0.9.}
\label{Fig:14}
\end{figure}
From Eqs. (5$\cdot$15) and (5$\cdot$16), the logarithmic term proportional to the color factor $C_AT_R$ is given by

\begin{eqnarray}& &2C_AT_R\frac{1}{z_2y_2}\log{z_1}-C_AT_R\frac{1}{z_2y_2}\log\frac{z_1z_2}{z_3}\nonumber \\&=&2C_AT_R\frac{1}{z_2y_2}\log\bigl(\frac{z_1z_3}{z_2}\bigr)^{1/2}.\end{eqnarray}
This term Eq. (5$\cdot$17) can be absorbed into the kinematical boundary of the phase space for two-body decay as follows:

\begin{eqnarray}&&\int^{t}_{M_0^2}\frac{dt_{13}}{t_{13}}2C_AT_R\frac{1}{z_2y_2}+2C_AT_R\frac{1}{z_2y_2}\log\bigl(\frac{z_1z_3}{z_2}\bigr)^{1/2} \nonumber \\&=&{}\int^{t(\frac{z_1z_3}{z_2})^{1/2}}_{M_0^2}\frac{dt_{13}}{t_{13}}2C_AT_R\frac{1}{z_2y_2}.\end{eqnarray}
After the absorption of the most singular term for the soft anti-quark region into the phase space of the two-body decay, the restriction on the phase space reduces to $t_{13}<t(z_1z_3/z_2)^{1/2}$. 
 
 In Fig. 15, since the contribution of the logarithmic term is small in the soft anti-quark region, the positivity of the three-body decay function for Process (G2) cannot be recovered by the subtraction of the effect of Eq. (5$\cdot$17). We need other modifications. As explained below, by the use of Eq. (5$\cdot$11), the logarithmic terms and the convolution terms of the non-logarithmic terms for the ladder diagrams $L^{(1)}$ and  $L^{(3)}$ can be absorbed into the kinematical boundaries of the two-body decays.  
 
 As explained above, we examine the contribution of the three-body decay function for the Process (G2) modified by Eqs. (5$\cdot$11) and (5$\cdot$18). Here, the ratio $R_{qg\bar{q}}$ is defined as

\begin{eqnarray}R_{qg\bar{q}}=\frac{L_L^{(1)}\log{y_1}+L_L^{(3)}\log{y_3}+(L_N^{(1)}+L_N^{(3)})(CV)}{V_{LL}}.\end{eqnarray}
Using Eq. (5$\cdot$19), the modified three-body decay function is given as
 
\begin{eqnarray}W^{\rm{New}}_{qg\bar{q}}=R_1-R_{qg\bar{q}}-\frac{C_AT_R}{(z_2y_2)V_{LL}}\log\bigl(\frac{z_1z_3}{z_2}\bigr),\end{eqnarray}
where $R_{qg\bar{q}}$ is given by Eq. (5$\cdot$19) and $R_1$ is given by Eq. (4$\cdot$1). In Fig. 15, the contribution of Eq. (5$\cdot$20) is positive for small $z_3$. Equation (5$\cdot$20) represents the three-body decay function modified by the new restriction resulting from the kinematical boundary of the phase space for the two-body decay. The function in Eq. (5$\cdot$20) can be applied for the region satisfying $z_2{\ge}z_3$. Thus, the new restriction due to the kinematical boundary of the phase space for the two-body decay can be applied to the region in which the contribution of Eq. (5$\cdot$12a) is negative.

\begin{figure}
\epsfxsize=10cm
\centerline{\epsfbox{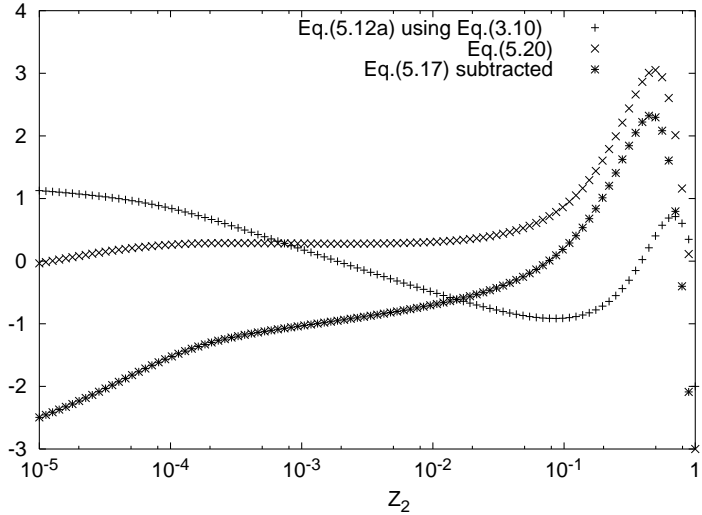}}
\caption{The $z_2$ dependence of Eq. (5$\cdot$12a) using Eq. (3$\cdot$10), that of Eq. (5$\cdot$20), and  that of contribution of the three-body decay function with Eq. (5$\cdot$17) subtracted at $z_3=10^{-4}$ for Process (G2).}
\label{Fig:15}
\end{figure}

\subsection{Modified non-logarithmic term for Process (Q1)}
  As shown in Fig. 4, in case that the gluon momentum fraction is hard in the soft quark region for Process (Q1), the effect of the soft gluon is small. Also, the contribution of the non-logarithmic term $V_N$ gives an important contribution in the soft quark region. We should evaluate the contribution of the NLL order for the parton shower model using the corrected method. 
 
 Following the approach used in Ref.~\citen{rf:4} and employing Eq. (5$\cdot$11), we adopt a modified  non-logarithmic term and obtain a positive value in the soft quark region. The non-logarithmic terms of the ladder diagrams for Process (Q1) are given as follows:

\begin{eqnarray}L_N^{(1)}&=&C_F^2\Bigl[-P(y_1)P\bigl(\frac{z_3}{y_1}\bigr)\frac{1}{y_1}+\frac{z_3}{y_1^2}\Bigr],\\L_N^{(2)}&=&{\hspace{5mm}}(z_1{\leftrightarrow}z_2){\hspace{5mm}}{\rm{in}} {\hspace{5mm}}L_N^{(1)},\\L_N^{(3)}&=&2C_FC_A\Bigl[-P(z_3)K\bigl(\frac{z_1}{y_3}\bigr)\frac{1}{y_3}+z_3\frac{(z_1-z_2)^2}{y_3^4}\Bigr].\end{eqnarray}

\begin{figure}
\epsfxsize=10cm
\centerline{\epsfbox{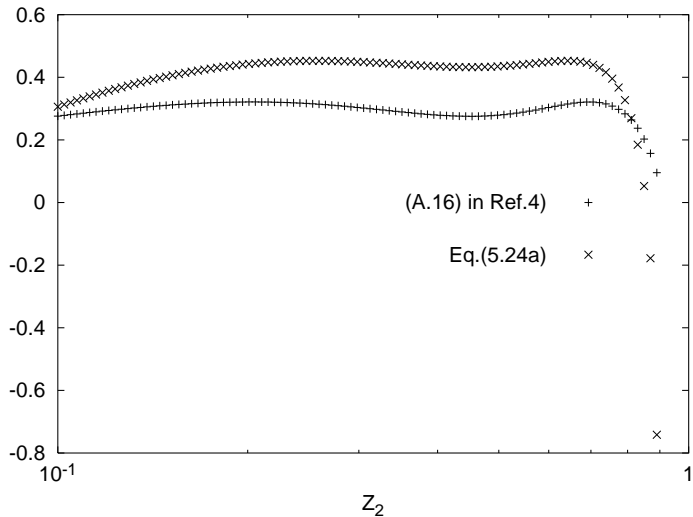}}
\caption{The $z_2$ dependences of Eqs. (5$\cdot$24a) and (A16a) taken from Ref. 4)   with $z_3=10^{-1}$ for Process (Q1).}
\label{Fig:16}
\end{figure}

Hence, the convolution terms of the Altarelli-Parisi splitting functions\cite{rf:11} in Eqs. (5$\cdot$21)--(5$\cdot$23) can be absorbed into the kinematical boundaries of the phase spaces of the two-body decays.\cite{rf:4} The modified three-body decay function is defined as

\begin{subequations}\begin{equation}V^{sq}_{qGG}=\frac{V_{NLL}-\displaystyle\sum^3_{i=1}L_N^{(i)}(CV)}{V_{LL}},\end{equation}
where 
\begin{equation}\sum^3_{i=1}L_N^{(i)}(CV)=-C_F^2P(y_1)P\bigl(\frac{z_3}{y_1}\bigr)\frac{1}{y_1}-C_F^2P(y_2)P\bigl(\frac{z_3}{y_2}\bigr)\frac{1}{y_2}-2C_FC_AP(z_3)K\bigl(\frac{z_1}{y_3}\bigr)\frac{1}{y_3}.\end{equation}\end{subequations}
As shown in Fig.~16, the contribution of the modified three-body decay function [Eq. (5$\cdot$24a)] is positive in the soft quark region. Although the correction of NLL order is larger than that reported in Ref.~\citen{rf:4}, the dominant contribution corresponding to the effect of the soft quark can be subtracted exactly.

\section{Summary}
 We have calculated three-body decay functions using the time-like jet calculus. Although most of the results agree with previous results obtained by two groups, some of the calculated results in this paper are different. The results we found for Processes (G1) and (Q2) agree with those reported in Refs.~\citen{rf:4} and \citen{rf:5}. Also, the result for Process (Q1) obtained in this paper is consistent with that of Ref.~\citen{rf:4}. However, the results we found for $I_L^{(1)}$ and $I_L^{(3)}$ of Process (G2) are not equivalent to those given in Refs.~\citen{rf:4} and \citen{rf:5}. 
 
 We also studied the properties of the three-body decay functions. First, we examined the gluon momentum fraction $z_2$ dependences of the ratios $R_1$ and $R_2$. We found that the contribution of the logarithmic term $V_L$ is dominant in most regions, with the exceptions of the soft quark region for Process (Q1) and the soft anti-quark region for Process (G2). 
 
 Using numerical results, we examined the structure of the strong cancellation for the non-logarithmic term $V_N$. Similarly to the situation considered in Ref.~\citen{rf:7}, we found that the contributions of the non-logarithmic terms for the interference diagrams are cancelled by those for the ladder diagrams and the other interference diagrams.
 
 We also studied the properties of the three-body decay functions in the soft parton regions. In the soft quark region for Process (Q1) and the soft anti-quark region for Process (G2), the contribution of the logarithmic term $V_L$ is small, because the contributions of the ladder diagrams and the interference diagrams cancel. 
 
 Because some of our results for the three-body decay functions differ from the results given in Refs.~\citen{rf:4} and \citen{rf:5}, the contribution of the modified three-body decay function for Process (G2) derived in Ref.~\citen{rf:4} was examined. We also corrected some of the results for the convolution terms of the non-logarithmic terms absorbed into the kinematical boundaries of the two-body decays given in Ref.~\citen{rf:4}. It was found that using the new results, the contribution of the modified three-body decay function used in Ref.~\citen{rf:4} is negative in some regions. Therefore, we suggested a new restriction introduced by the kinematical boundary of the phase space for the two-body decay with modifications of the non-logarithmic terms. The result of this change is that the contribution of the three-body decay function for Process (G2) is positive in the region satisfying $z_3{\ll}z_2{\ll}z_1$.
 
 The contribution of NLL order in the soft quark region for Process (Q1) was examined. The non-logarithmic term $V_N$ corresponding to the dominant contribution in the soft quark region was subtracted as the effect of the soft gluon in NLLJET. Considering this, we suggested that some of the non-logarithmic terms for the ladder diagrams should be absorbed into the kinematical boundaries of the two-body decays. 
 
 To summarize, we derived new three-body decay functions for time-like branching and carried out a numerical treatment of the three-body decay functions in the soft parton regions for the parton shower model. In a future paper, we will present the NLL parton shower model improved by the new results and the methods presented in this paper.
 
\section*{Acknowledgements}
 The author would like to express his thanks to Professor H. Tanaka. He is also grateful for several useful discussions with Professor T. Munehisa and members of Group of Theoretical Physics, Rikkyo University. This work was supported in part by Rikkyo University Research Center for Measurement in Advanced Science.

\appendix
\section{Kinematics}
 In this appendix, we present the details of the calculations for the decay functions of Types [L1] and [I3]. The calculations for the other types  can be obtained by exchanging the momenta $l_i$ in Types [L1] and [I3]. 

\subsection{Calculation of Type $[$L1$]$ $($Ladder diagram$)$}
The combination $X=z_1z_2$ is used to determine the decay function for Type [L1].
The constraint in the $\delta(l_3^2)$ function is written   
\begin{eqnarray}0=l_3^2=(q-l_1-l_2)^2=z_3t+\bigl(-\frac{z_3}{z_1y_1}\bigr){\vec{l}}\,^2_{1T}-\frac{y_1}{z_2}{\vec{p}}\,^2_{2T}.\end{eqnarray}
Here, $\vec{l}_{2T}$ has been replaced by introducing the new vectors ${\vec{p}}_{2T}$ and $\vec{l}_{1T}$ according to

\begin{eqnarray}\vec{l}_{2T}=-\frac{z_2}{y_1}\vec{l}_{1T}+\vec{p}_{2T},\end{eqnarray}
with $y_i=1-z_i\;\;(i=1,2,3)$.
 
From Eq. (A$\cdot$1), ${\vec{p}}\,_{2T}^2$ is given by

\begin{eqnarray}{\vec{p}}\,_{2T}^2=\frac{z_2z_3}{y_1}\bigl(t-\frac{{\vec{l}}\,_{1T}^2}{z_1y_1}\bigr).\end{eqnarray}
${\vec{l}}\,^2_{1T}$ is written as the invariant

\begin{eqnarray}{\vec{l}}\,^2_{1T}=z_1y_1t-z_1t_{23}.\end{eqnarray}
By use of Eq. (A$\cdot$4), ${\vec{p}}\,^2_{2T}$ is given by

\begin{eqnarray}{\vec{p}}\,^2_{2T}=\frac{z_2z_3t_{23}}{y_1^2}.\end{eqnarray}
It is easy to calculate the term ${\P}M$ in Eq. (2$\cdot$9) using the constraint represented by the delta function. The general form of the term ${\P}M$ is given by
 
\begin{eqnarray}{\P}M=\frac{F_1t^2+F_2t_{12}^2+F_3t_{23}^2+F_4t_{12}t_{23}+F_5tt_{12}+F_6tt_{23}}{t_{23}^2},\end{eqnarray}
where the $F_i$ are functions of the $z_i$. The variable $t_{12}$ depends on the azimuthal angle $\phi$. The variable $t_{12}$ is replaced by Eq. (A$\cdot$2):

\begin{eqnarray}t_{12}=(l_1+l_2)^2=\bigl(\frac{z_2}{z_1}+\frac{z_1z_2}{y_1^2}+\frac{2z_2}{y_1}\bigr)\vec{l}\,^2_{1T}+\frac{z_1}{z_2}
\vec{p}\,^2_{2T}-\frac{2}{y_1}|\vec{p}_{2T}||\vec{l}_{1T}|\cos\phi.\nonumber \\\end{eqnarray}
Here, the term with dependence on the azimuthal angle in Eq. (A$\cdot$7) can be isolated. It follows from Eqs. (A$\cdot$4) and (A$\cdot$5) that $t_{12}$ and $t_{12}^2$ in the numerator satisfy
 
\begin{eqnarray}t_{12}&=& \frac{z_2}{y_1}t+\frac{z_1z_3-z_2}{y_1^2}t_{23}-\frac{2}{y_1}|{\vec{p}}_{2T}||{\vec{l}}_{1T}|\cos\phi,\\
t_{12}^2&{=}& \bigl[\frac{z_2}{y_1}t+\frac{z_1z_3-z_2}{y_1^2}t_{23}\bigr]^2+4\frac{z_1z_2z_3}{y_1^4}\bigl(y_1t-t_{23}\bigr)t_{23}\cos^2\phi.\end{eqnarray}
After replacement of the invariant in Eq. (A$\cdot$6), we obtain 
   
\begin{eqnarray}{\mbf{P}}^{(i)}M=\frac{G_1t^2+G_2t_{23}^2+G_3tt_{23}+(G_4t_{23}^2+G_5t_{23}t)\cos^2\phi+(G_6t_{23}+G_7t)\cos\phi}{t_{23}^2},\nonumber \\\end{eqnarray}
where the $G_i$ are functions of the $z_i$. Although $G_6$ and $G_7$, which are proportional to $\cos\phi$, are also functions of ${\vec{p}}_{2T}$ and ${\vec{l}}_{1T}$, they drop out through the azimuthal integration. Since the mass singularity is of logarithmic order, there is no singularity term of higher than logarithmic order. Thus, we obtain the condition

\begin{eqnarray}G_1=0.\end{eqnarray}
 
 Next, using Eqs. (A$\cdot$4) and (A$\cdot$5), the phase space volume element is given by
\begin{eqnarray}d\tilde{\Gamma}=\frac{\pi}{2}z_1dt_{23}d{\vec{p}}\,^2_{2T}d\phi\delta(l_3^2).\end{eqnarray}
The integration of the $\delta$ function is trivial. The kinematical boundary of the integration over the variable $t_{23}$ is determined from the conditions ${\vec{p}}\,^2_{2T}>0$ and ${\vec{l}}\,^2_{1T}>0$. We obtain 

\begin{eqnarray}0<t_{23}<y_1t.\end{eqnarray}
Therefore, the formula for the phase space volume element in the jet calculus is given by

\begin{eqnarray}d\tilde{\Gamma}=\frac{z_1z_2\pi^2}{y_1}\int^{y_1t}_{M_0^2}dt_{23}\frac{d\phi}{2\pi}.\end{eqnarray}

 We integrate the term ${\P}M$ over the azimuthal angle $\phi$ to obtain
 
\begin{eqnarray}\int\frac{d\phi}{2\pi}{\P}M=\frac{H_2t_{23}^2+H_3tt_{23}}{t_{23}^2},\end{eqnarray}
where the $H_i$ are functions of the $z_i$. After the integration over the azimuthal angle, the terms proportional to $\cos^2\phi$ in Eq. (A$\cdot$10) are absorbed into the other terms. In order to obtain the decay function $J$ in Eq. (2$\cdot$19), integration over the variable $t_{23}$ is performed. We then obtain
 
\begin{eqnarray}J&=&\frac{1}{4y_1}\int^{y_1t}_{M_0^2}\bigl[H_3\frac{dt_{23}}{t_{23}}+H_2\frac{dt_{23}}{t}\bigr]
\nonumber \\&=&{}\frac{1}{4y_1}\int^{t}_{M_0^2}H_3\frac{dt_{23}}{t_{23}}+\frac{1}{4y_1}H_3\log{y_1}\nonumber \\
& &{}+\frac{1}{4}H_2+O\bigl(\frac{M_0^2}{t}\bigr).\end{eqnarray}
The first term and the other terms in Eq. (A$\cdot$16) correspond to the contributions of LL order and NLL order, respectively. The contribution of NLL order is written 

\begin{eqnarray}\frac{1}{4y_1}H_3\log{y_1}+\frac{1}{4}H_2=L_L^{(1)}\log{y_1}+L_N^{(1)},\end{eqnarray}
with

\begin{eqnarray}L_L^{(1)}=\frac{1}{4y_1}H_3,L_N^{(1)}=\frac{1}{4}H_2.\end{eqnarray}
Since the two-body decay function includes $O(M_0^2/t)$, we do not need this term in the three-body decay function.

\hspace{10mm}
\subsection{Calculation of Type $[$I3$]$ $($Interference diagram$)$}
 The decay functions for the interference diagrams are free from mass singularities. Here we construct the condition for these decay function to possess no mass singularity.
 
We start by constructing the phase space volume element as follows:
\begin{eqnarray}d\tilde{\Gamma}=\frac{\pi}{2}d({\vec{l}}\,^2_{1T})d({\vec{l}}\,^2_{2T})d\phi\delta(l_3^2).\end{eqnarray}
The new vectors ${\vec{l}}_{1T}$ and ${\vec{l}}_{2T}$ are defined as

\begin{eqnarray}{\vec{h}}_{1T}={\vec{l}}_{1T}, -\frac{z_2}{y_1}{\vec{h}}_{1T}+\frac{z_3}{z_1y_1}{\vec{h}}_{2T}={\vec{l}}_{2T}.
\end{eqnarray}
We find the phase space volume element from Eq. (A$\cdot$20):

\begin{eqnarray}d\tilde{\Gamma}=\frac{\pi}{2}\bigl(\frac{z_3}{z_1y_1}\bigr)^2d({\vec{h}}_{1T}^2)d{\phi}d({\vec{h}}_{2T}^2)\delta({l_3^2}).\end{eqnarray}
The constraint represented by the $\delta$ function is expressed as

\begin{eqnarray}0=l_3^2=z_3t-\frac{z_3}{z_1y_1}{\vec{h}}_{1T}^2-\frac{z_3^2}{z_2z_1^2y_1}{\vec{h}}_{2T}^2.\end{eqnarray} 

 Next, we calculate the term ${\P}M$ in Eq. (2$\cdot$9). After eliminating $t_{12}$ using Eq. (2$\cdot$8), the general formula for the term ${\P}M$ becomes

\begin{eqnarray}{\P}M=\frac{F_1t^2+F_2t_{13}t_{23}+F_3tt_{13}+F_4tt_{23}+F_5t_{13}^2+F_6t_{23}^2}{t_{23}t_{13}}.\end{eqnarray} 
We absorb the last two terms into the other terms in Eq. (A$\cdot$23). The last two terms are given by
\begin{eqnarray}K_1={\int}d\tilde{\Gamma}\frac{t_{13}^2}{t_{13}t_{23}}={\int}d\tilde{\Gamma}\frac{t_{13}}{t_{23}}
\end{eqnarray}
and
\begin{eqnarray}K_2={\int}d\tilde{\Gamma}\frac{t_{23}^2}{t_{13}t_{23}}={\int}d\tilde{\Gamma}\frac{{t_{23}}}{t_{13}}.\end{eqnarray}
After $z_2$ in Eq. (A$\cdot$8) is replaced by $z_3$, the invariant $t_{13}$ is given by
 
\begin{eqnarray}t_{13}=\frac{z_3}{y_1}t+\frac{z_1z_2-z_3}{y_1^2}t_{23}-\frac{2}{y_1}|{\vec{p}}_{3T}||{\vec{l}}_{1T}|\cos\phi.\end{eqnarray}
Similarly, $z_1$ in Eq. (A$\cdot$26) can be replaced by $z_2$. The invariant $t_{23}$ is given by
\begin{eqnarray}t_{23}=\frac{z_3}{y_2}t+\frac{z_1z_2-z_3}{y_2^2}t_{13}-\frac{2}{y_2}|{\vec{p}}_{3T}||{\vec{l}}_{2T}|\cos\phi.\end{eqnarray}
Using Eqs. (A$\cdot$26) and (A$\cdot$27), we obtain the formula for the term ${\P}M$ as
 
 \begin{eqnarray}{\P}M={\int}d\tilde{\Gamma}\frac{G_1t_{13}t_{23}+G_2t^2+G_3tt_{13}+G_4t_{23}t+(G_5t_{23}+G_6t_{13})\cos\phi}{t_{13}t_{23}},\nonumber \\
 \end{eqnarray} 
where the $G_i$ are functions of the $z_i$. $G_5$ and $G_6$ are functions of ${\vec{p}}_{3T}$ and ${\vec{l}}_{1T},{\vec{l}}_{2T}$. The terms proportional to $\cos\phi$ can be dropped due to the azimuthal integration.
 
 The interference term must be free from mass singularities, which may occur at $t_{13}=0$ or $t_{23}=0$. The term $G_1t_{13}t_{23}$ in Eq.~(A$\cdot$28) has no mass singularity. The terms proportional to $G_2,G_3$ and $G_4$ must also be free of mass singularities. Here we present a non-trivial combination of the functions $G_i$ which results in no mass singularity. The invariants $t_{13}$ and $t_{23}$ are written 
 
 \begin{eqnarray}t_{23}=\frac{z_3}{z_1^2z_2}{\vec{h}}_{2T}^2,\end{eqnarray}
 \begin{eqnarray}t_{13}=\frac{z_3}{z_1y_1^2}\bigl({\vec{h}}_{1T}+{\vec{h}}_{2T}\bigr)^2.\end{eqnarray}
Using Eqs. (A$\cdot$4), (A$\cdot$29) and  (A$\cdot$30), the formula for the azimuthal integration satisfying the condition that there be no mass singularity is given by
 
\begin{eqnarray}& &\frac{-2y_1z_1^2z_2}{z_3^2}\int^{2\pi}_0{d\phi}\frac{{\vec{h}}_{2T}^2+{\vec{h}}_{1T}{\vec{h}}_{2T}}{{\vec{h}}_{2T}^2({\vec{h}}_{1T}+{\vec{h}}_{2T})^2}\nonumber \\&=&{}-\frac{2y_1z_1^2z_2}{z_3^2}\theta({\vec{h}}_{2T}^2-{\vec{h}}_{1T}^2)\nonumber \\&=&{}\int^{2\pi}_0{d\phi}\frac{1}{t_{13}t_{23}}\bigl(t-\frac{y_1}{z_3}t_{13}-\frac{y_2}{z_3}t_{23}\bigr),\end{eqnarray}
where $\theta$ is the step function. The relation between Eq. (A$\cdot$28) and Eq. (A$\cdot$31) is given by

\begin{eqnarray}G_3=-\frac{y_1}{z_3}G_2,\hspace{5mm}G_4=-\frac{y_2}{z_3}G_2.\end{eqnarray}
Substituting this combination into Eq. (A$\cdot$28), we have

\begin{eqnarray}{\int}d\tilde{\Gamma}\Bigl[G_1+G_2\frac{t}{t_{13}t_{23}}\bigl(t-\frac{y_1}{z_3}t_{13}-\frac{y_2}{z_3}t_{23}\bigr)\Bigl].\end{eqnarray} 
Then, the latter integrations can be performed:

\begin{eqnarray}K_0&=& {\int}d\tilde{\Gamma}\frac{t_{13}t_{23}}{t_{13}t_{23}},\nonumber \\K_3&=& 
{\int}d\tilde{\Gamma}\frac{t}{t_{13}t_{23}}\bigl(t-\frac{y_1}{z_3}t_{13}-\frac{y_2}{z_3}t_{23}\bigr).\end{eqnarray}
The ${\P}M$ term in $K_0$ is independent of the azimuthal angle, while the ${\P}M$ term in $K_3$ depends on the azimuthal angle. To begin, we integrate the phase space in $K_0$. Substituting Eq. (A$\cdot$29) into Eq. (A$\cdot$21) and integrating over ${\vec{h}}_{1T}^2$, we have
 
 \begin{eqnarray}d\tilde{\Gamma}=\frac{\pi^2z_1z_2}{y_1}{\int_{0}^{y_1t}}dt_{23}\frac{d\phi}{2\pi}.\end{eqnarray}
The region of integration over the invariant $t_{23}$ is the same as that in Eq. (A$\cdot$13). We obtain

\begin{eqnarray}K_0=\pi^2z_1z_2t.\end{eqnarray}
Next, we compute $K_3$. The result obtained from integrating $K_3$ over the azimuthal angle corresponds to the second term in Eq. (A$\cdot$31). Due to the constraint introduced by the $\delta$ function and the step function [Eqs. (A$\cdot$22) and  (A$\cdot$31)], the kinematical boundary of $K_3$ is given by

\begin{eqnarray}{\vec{h}}_{1T}^2{\geq}0{\to}t_{23}<y_1t,\\{\vec{h}}_{1T}^2{\leq}{\vec{h}}_{2T}^2{\to}\frac{z_3}{y_2}t{\leq}t_{23}.\end{eqnarray} 
Therefore, the result for $K_3$ is   

\begin{eqnarray}K_3=-\frac{2\pi^2z_1z_2}{z_3}t\int^{y_1t}_{\frac{z_3}{y_2}t}\frac{dt_{23}}{t_{23}}
=-\frac{2\pi^2z_1z_2}{z_3}t\log\frac{y_1y_2}{z_3}.\end{eqnarray}
Thus, the decay function of Type [I3] is given by

\begin{eqnarray}J&=& \frac{1}{4\pi^2z_1z_2t}\bigl(K_0G_1+K_3G_2\bigr)\nonumber \\&=& \frac{1}{4}G_1-\frac{1}{2z_3}G_2\log
\frac{y_1y_2}{z_3},\end{eqnarray}
with

\begin{eqnarray}I^{(3)}_L=-\frac{G_2}{2z_3},I^{(3)}_N=\frac{G_1}{4}.\end{eqnarray}

\end{document}